\documentclass{elsart}

\newcounter{bla}

\usepackage{graphicx}

\def \beq{\begin{equation}}
\def \eeq{\end{equation}}
\def \beqar{\begin{eqnarray}}
\def \eeqar{\end{eqnarray}}

\begin{document}
\begin{frontmatter}

\title{Understanding and characterizing  nestedness in mutualistic bipartite networks}

\author[a,b]{Enrique Burgos},
\author[b]{Horacio Ceva},
\author[c]{Laura Hern\'andez\thanksref{author}}
\author[d]{R.P.J. Perazzo}

\thanks[author]{Corresponding author\\
 e-mail: Laura.Hernandez@u-cergy.fr}

\address[a] {Consejo Nacional de Investigaciones Cient{\'\i }ficas y T\'ecnicas,\\
Avenida Rivadavia 1917, C1033AAJ, Buenos Aires, Argentina}
\address[b]{Departamento de F{\'{\i}}sica, Comisi{\'o}n Nacional de Energ{\'\i }a
At{\'o}mica,\\ Avenida del Libertador 8250, 1429 Buenos Aires, Argentina}
\address[c]{Laboratoire de Physique Th\'eorique et Mod\'elisation; \\
UMR CNRS, Universit\'e de Cergy-Pontoise, \\
2 Avenue Adolphe Chauvin, 95302, Cergy-Pontoise Cedex France}
\address[d] {Departamento de Investigaci\'{o}n y Desarrollo, Instituto Tecnol\'{o}gico de
Buenos Aires \\ Avenida E. Madero 399, Buenos Aires, Argentina}

\begin{abstract}
  %Type your abstract here.
In this work we present a dynamical model that succesfully describes the organization of mutualistic ecological systems.
The main characteristic of these systems is the nested structure of the bipartite adjacency matrix describing their
interactions.  We introduce a
nestedness coefficient, as an alternative to the Atmar and Patterson temperature, commonly used to measure the nestedness
degree of the network. This coefficient has the advantage of being based on the robustness of the ecological system
and it is not only describing  the ordering of the bipartite matrix but it is also able to tell the difference, if any,
between the degree of organization of each guild.

\begin{flushleft}
  %Insert your suggested PACS number here
%PACS code one; PACS code two; PACS code three; Etc.
\end{flushleft}

\begin{keyword}
ecological mutualists systems, bipartite complex graphs, nested networks

  % Please give some freely chosen keywords that we can use in a
  % cumulative keyword index.
\end{keyword}

\end{abstract}

\end{frontmatter}

\section{Introduction\label{seccion_Introduction}}
 Bipartite multigraphs are useful tools to describe systems with two
different kinds of agents and such that the only allowed interactions
involve agents of different kinds.
 
A very interesting example is an ecological system consisting of 
 two groups of species, usually
animals and plants, that interact to fulfill essential biological functions such as feeding or
reproduction. This is the case of systems involving plants and animals that feed from the fruits
and disperse their seeds (\textit{seed dispersal networks}). Another example is that
of insects that feed from the nectar of flowers while pollinating them in the process
(\textit{pollination networks}).

Bipartite networks can also be found in social systems. Examples of this type involve
the actors and movies in which they participate \cite{W&S} or the
boards of directors of large companies and their members \cite{Newman}.

The interaction pattern of a bipartite network can be coded as an
adjacency matrix in which rows and columns are labeled respectively by the plant and animal
species involved in the network. Its elements $K_{p,a} \in \{0,1\}$ represent respectively
the absence  or the presence of an interaction between the plant species $p$ and the
animal species $a$. In what follows we drop the term species specifying that
when mentioning plant or animals we are not referring to the behavior of separate
individuals but to all the members of a species.

It has been found that bipartite networks describing the interactions in natural ecological
 systems  have a very special structure called {\em nestedness}~\cite{AyP}~\cite{muchas}.
In a nested network the nodes of both types can be ordered by
decreasing degree in such a way that all the links of a given species are
subset of the links of the preceding one. 
This organization is such that the \textit{generalists} of both types
of guilds (i.e. those nodes that interact with a great number of nodes of the other guild)
tend to interact among them while there are very few contacts among \textit{specialists} (i.e.
nodes that interact with very few ones of the other guild).
When the rows and columns of the two interacting guilds are ordered in decreasing degree most of
 the contacts lie under a curve \cite{AyP},
\cite{nosotros1} called \textit{isocline of perfect nestedness}(IPN).
 All these features indicate that
these networks are far from being a random collection of interacting species, displaying
instead a high degree of internal organization.

This particular structure can be thought of  as  the
outcome of a self-organization process.  In 
~\cite{nosotros1}~\cite{nosotros2}   the Self-organizing Network Model (SNM) has been introduced.
In this model the nodes progressively redefine their links
obeying to a purely local rule that does not depend upon any global feature of the
network.

Here we briefly describe the SNM and we refer
the reader to previously cited references  for the details of the original model as well as
 for the comparison of its results with the empirical observations.

\section{Description of the Model and Results}

The SNM is a computer model that starts from a random adjacency matrix in which the number of
plants, animals and of the contacts between them are arbitrarily fixed provided
that there are no species left without links with the other guild. Starting from this initial
configuration plants and animals iteratively redefine their contacts by reallocating
the contacts of the adjacency matrix. 
This reallocation obeys to some assumed  {\em contact preference rule} (CPR). In the  following we
consider a CPR that indicates that the agents of either guild prefer to set contacts
with a species of the opposite guild having a greater (or lesser) number of contacts.

In each iteration of the SNM algorithm first a row and 
next a column
are chosen at random. Once a row (column) has been chosen, its contacts are reallocated with
probability $P_r$ ($P_c$). Reallocation consists in choosing at random a 1 and a 0 of the same 
row (column) and swapping them according to a previously selected CPR.
 For instance in the version SNM-I (SNM-II) the CPR indicates that the proposed swapping actually
takes place if the degree of the new partner is higher (lower) than the degree of the initial one.

In case that, upon swapping, a row or a column would be left with no links, the
reallocation is not produced. This rule prevents the elimination of a node of the system as
a consequence of being left without interactions.

At first sight the CPR of SNM-I looks very similar to the {\em preferential attatchment rule} but
two important differences must be underlined: while preferential attatchment deals with a growing  network, here we work with a fixed number of species. Moreover, unlike preferential attatchment
this CPR is local as it doesn't depend on the whole degree distribution of the network.

\begin{figure}
%\hspace{-1.5cm}\includegraphics[width=10cm]{fig1.eps}
\includegraphics[width=12cm]{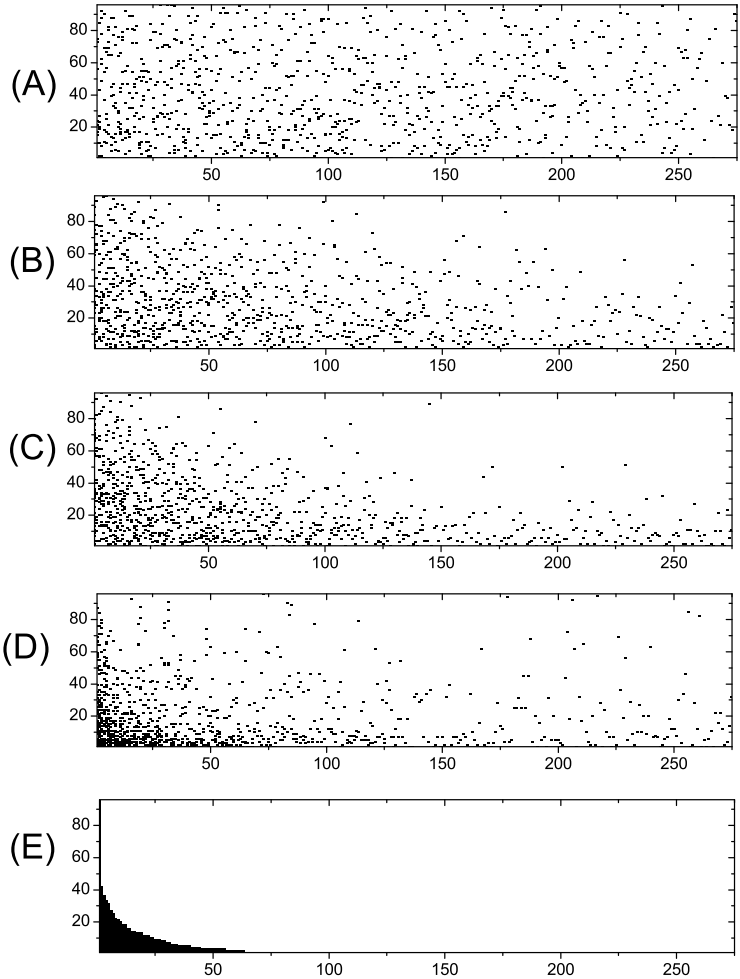}
\caption{\footnotesize Adjacency matrices of a bipartite network corresponding to the size and
density of contacts of the real system studied in~\cite{clements} . Each
contact is shown as a black pixel. Panels display the adjacency matrix obtained with the
SNM using Strategy (I) for both rows and columns at different times t, measured in number of iteration steps, (A):initial random matrix, (B): after t=1000 SNM iterations, (C): at this 
iteration time, t=2175,  the simulated system has the same $T_{AP}$ 
value that in the real system (D), $T_{AP} \approx 2.41$, (E): perfectly nested system, 
corresponds to 100000 iterations of the SNM  .} 
\label{contactos}
\end{figure}

SNM-I always leads to a perfectly nested
pattern, no matter the relative updating probability of rows and columns~\cite{nosotros3}.
Figure~\ref{contactos} shows different stages of the evolution of the SNM.
 The initial random matrix has been chosen so as to have the same number
 of rows, columns and density of contacts as the real system studied by Clements and Long~\cite{clements}. 
Panel D corresponds to that real system and though it is clear that the ordering process has 
already started (see panel A), the emergent state is very far from the perfectly nested state shown
in E.
Hence, in order to reproduce 
the real observed networks,  a {\em stopping criterium} for the iteration process of
 the SNM algorithm is needed. Therefore one needs a measure of the  
the degree of nestedness. A widespread 
parameter doing so is the so-called "Temperature" introduced by Atmar and Patterson~\cite{AyP},
$T_{AP}$.  
In panel C  one can see for comparison the result of the SNM iterated till its $T_{AP}$ approaches
the value obtained for the real system.

The $T_{AP}$   has been largely used in the biogeography and ecology communities~\cite{calculator}, 
as a global measurement of nestedness. 
In principle, a  "high temperature" indicates a disordered lattice while a low one indicates
 that the system is  nested. Nevertheless one can easily verify that this parameter is 
highly dependent on the density of contacts and also on the size of the  adjacency matrix. 
Table~\ref{TAP}  illustrates this dependence. The $T_{AP}$ values correspond to averages
over 200 random matrices
 having the same sizes, shapes and density of contacts, $\phi$,   found in known real systems.

\begin{table}
\begin{center}
\begin{tabular}{|c|c|c|c|c|}
\hline
Random adjacency matrix                 & $\phi$        & $T_{AP}$ &    $m$     & $n$ \\ \hline
simil Clements  &  3.4\%        & 15       &    96      & 275  \\ \hline
simil Robertson &  2.3\%        & 12.7     &    456     & 1428 \\ \hline
simil Kato      &  1.9\%        & 8.7      &    93      & 679  \\ \hline

\end{tabular}
%\end{center}

\vspace{1cm}

\caption{\footnotesize $T_{AP}$ of random matrices. Dependence  on size, shape and density of contacts.
 Average over 200
 random matrices having the same size, shape and density of contacts as in references~\cite{clements}
~\cite{Robertson}~\cite{Kato}} 

\label{TAP}
\end{center}
\end{table}

The theoretical basis of $T_{AP}$ have been given in ~\cite{nosotros1} where it is
shown how it accounts for  the average departure of the matrix from the 
IPN curve.

An alternative parameter to measure the degree of nestedness can be obtained by studying the
 robustness of the system face to 
different types of perturbations or attacks. In~\cite{nosotros2} robustness coefficient was 
introduced and the theoretical values were obtained in limiting cases. If a given fraction of 
species of one guild dissapears (is attacked) some species of the other guild may remain without
interaction and hence become extinct.
Memmott et al~\cite{memmott} defined the Attack Tolerance Curve (ATC) as the  curve 
giving the fraction of surviving  species of one guild  as a
 function of the fraction of eliminated (attacked) species of the other guild. To fix ideas let's
 consider the
case where we have the pollinators in the {\em n}  columns and the  plants in the {\em m}
 rows of the 
bipartite matrix. Then the  ATC gives the fraction of surviving plants as a function of the 
fraction of
extinct animals, $S_p(f_a)$.

The derivation of the ATC from the adjacency matrix requires some additional assumptions. 
First, it has been assumed that the extinction of a plant (animal) species occurs after all its contacts have been removed. 
It is also necessary to specify the order in which the animal (plants) species are eliminated. 
The least biased choice is to assume that all species have the same probability of becoming extinct.
This is the {\em null model}. There are two alternative, highly schematic ways in which the species 
can be eliminated: starting  from the highest degree  animal (plant), noted $(+ \rightarrow -)$
 or the opposite scheme, starting from the lowest degree animal (plant), noted $(- \rightarrow +)$.
 These two schemes are equivalent to assume that species have a different probability of becoming
 extinct depending upon their number of contacts.
Finally, to build the ATC for the null model, it is necessary to produce a statistically significant result by averaging the calculations over several realizations. 
In each realization all the contacts of the same fraction of randomly selected column-species are set to 0, and then the fraction of row-species that become extinct is evaluated.
The resistence of the network to the different kinds of attacks depends on the organization of the
 network.

\begin{figure}

%\hspace{-1.3cm}\includegraphics[width=10cm]{fig2.eps}
\hspace{1.5cm}\includegraphics[width=10cm]{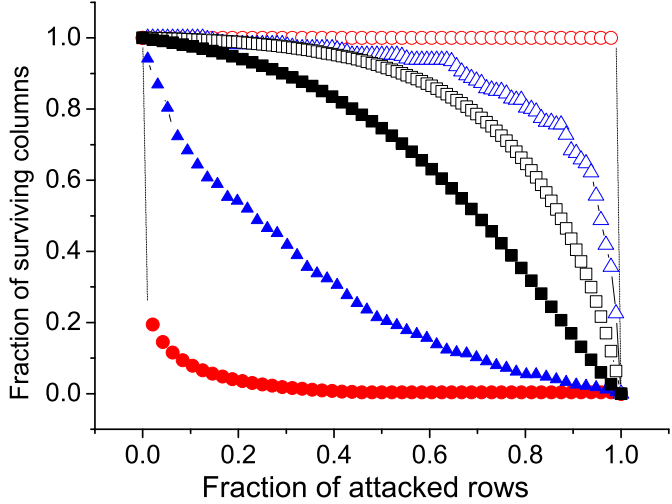}
\caption{\footnotesize  Attack tolerance curves for a matrix of the same size and 
density of contacts as the real system in~\cite{clements}. Open
symbols correspond to the attack strategy $(- \rightarrow +)$ and solid symbols
to the attack strategy $(+ \rightarrow -)$. Squares: average ATC over 200  random initial 
configurations, circles: SNM evolved nested state (average over 200 initial configurations),
 triangles: real system.}
 
\label{ATC}

\end{figure}

In~\cite{nosotros2} the robustness coefficient is defined as the area under the ATC curve.
 Hence it depends on the applied attack strategy.
In Figure~\ref{ATC}  one can see  that in the case of perfect nestedness, 
where all the contacts are under the IPN, the
attack strategy  $(- \rightarrow +)$ leads to an ATC which is a  discontinuous function, 
and $R^{- \rightarrow +}=1$. On the other hand,  when applying the attack 
strategy $(+ \rightarrow -)$ it is easy to see that the ATC is equivalent to  the IPN curve
 normalised to 1, leading to $R^{+ \rightarrow -}=\phi$.

For  random matrices, the two attack strategies lead  to ATCs having the same curvature. 
This is also
shown in Figure~\ref{ATC} where the results are obtained after averaging over 200 realizations
of random bipartite matrices.

Since the  adjacency  matrices corresponding to real 
systems are neither random nor perfectly nested, the corresponding ATCs lie 
between the perfectly nested  and the random one, for the two attack strategies respectively.

\begin{figure}
%\hspace{-1.5cm}\includegraphics[width=10cm]{fig3.eps}
\hspace{1.5cm}\includegraphics[width=10cm]{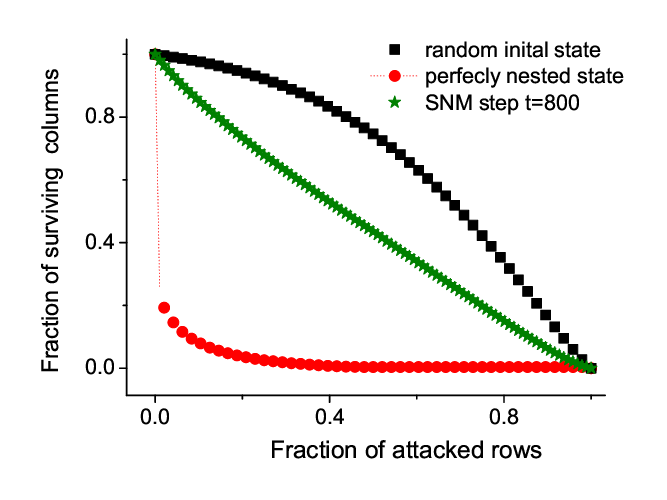}
\caption{\footnotesize  Attack tolerance curves for a matrix of the same size and
density of contacts as the real system of Clements and Long~\cite{clements},under the attack strategy $(+ \rightarrow -)$, for three different
stages of the evolution of the SNM algorithm. Circles:
average ATC over 200 SNM evolved perfectly nested networks , 
squares: average ATC over 200 random networks, 
stars: average ATC over 200 SNM evolved networks at an intermediate iteration time 
 of the SNM algorithm, here $t^* \approx 800$ steps.}

\label{curvature}
\end{figure}

It is interesting to notice that the ATC curves corresponding to the attack strategy $R^{+ \rightarrow -}$ change their curvature as the system evolves according
to the SNM algorithm. Figure~\ref{curvature} shows three ATCs corresponding to averages over 
200 random initial 
matrices having  the same caracteristics as the real system of
 Clements and Long~\cite{clements} for three different stages of its evolution, from random to perfect
 nestedness.
 Hence, the zero curvature of the ATC  may be taken as the indication that 
the network has lost its random character.
In the SNM,  the iteration step where the ATC has zero curvature,  can be interpreted as
 an ordering time $t^*$. In this exemple this is found at $t^*\approx 800$.

We can then define a %{\em nesting parameter}%
{\em nestedness coefficient}
 N as the normalised difference between 
the robustness coefficients corresponding to the two extreme attack strategies:
\begin{equation}
N=\frac {R^{- \rightarrow +}-R^{+ \rightarrow -}}{1-\phi}
\end{equation}

Notice that as $1-\phi$ is the area between the two ATC in the case of perfect nestedness, 
this coefficient is correctly bounded, $N=1$ for perfect nestedness and $N$ decreases for
 random matrices~\cite{antinested}.

\begin{figure}
%\hspace{-1.5cm} \includegraphics[width=9cm]{fig4.eps}
\hspace{1.5cm} \includegraphics[width=10cm]{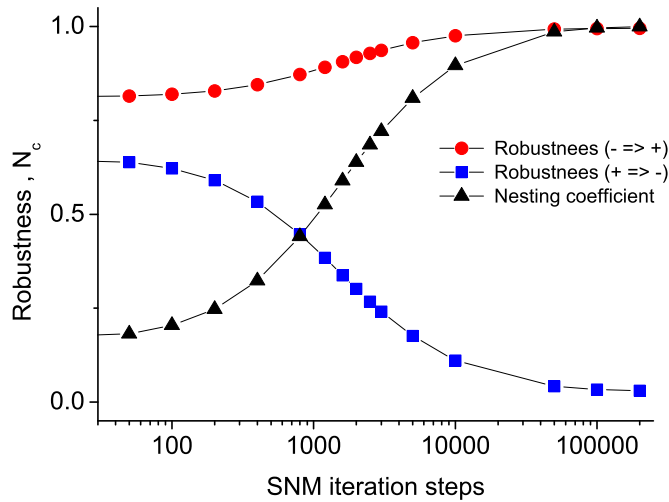}
\caption{\footnotesize Robustness and Nesting coefficient of the animals (columns)
 as a function of iteration time of the SNM algorithm, average over 200 initial random networks. 
Circles correspond to robustness under strategy  $R^{- \rightarrow +}$,  squares  correspond 
to robustness under strategy  $R^{+ \rightarrow -}$ and triangles correspond to the
 nestedness coefficient.}

\label{nescoef}
\end{figure}

In Figure~\ref{nescoef} we show the two robustness coefficients and the  nesting parameter 
 as a function of the iteration time of the SNM algorithm, starting from  random configurations 
and averaging over 200 samples. 
The inflexion point of the N(t) is also located at  $t^* \approx 800$,   
the  time of the SNM evolution  which leads to an  ATC  with  zero curvature,  
indicating again that the ordered phase is setting on.

The analysis we just described  assumed that plants (rows)  were attacked, leading to a 
coefficient 
measuring  the robustness of the animals (columns),  $N_c$.  Obviously the same analysis can
 be done 
assuming that the animals  are attacked, which  will lead to a coefficient measuring the 
robustness of plants,  $N_r$. 
In this way, the nestedness of the network is characterised by two parameters $N_r$ and $N_c$.
For real systems~\cite{nosotros1} it is  generally found that the number of animal species is higher 
that the number of plants species and in that case the SNM model always gives $N_c > N_r$. 
This is easy to understand in terms of the SNM algorithm. In
fact at a given stage of the SNM iteration, if the update probability is the same 
for rows and columns,  rows are chosen, in average, more often than columns, due to their
 relative smaller 
number. Choosing a row means that the update is done over the columns, so at a given 
iteration time
columns are more ordered than rows. This has been checked by simulating a square adjacency
matrix having  the same  total
number of sites and the same density as the real system studied in~\cite{clements}, in this 
case $N_r = N_c$ for all the SNM evolution. Surprisingly, the same 
result, $N_c > N_r$, is found for adjacency matrices corresponding to real 
networks\cite{clements}~\cite{Robertson}~\cite{Kato}. 
This is shown in table~\ref{T&N} where we give the   normalized Atmar-Patterson temperature, 
  ${\tilde T_{AP}}= T_{AP}/T_{AP,ran}$, with  $T_{AP,ran}$, the
temperature of the random matrix of the same dimensions and the same density of the
considered one. In this way, complete disorder gives ${\tilde T_{AP}}=1$

\begin{table}
\begin{center}
\begin{tabular}{|c|c|c|c|}
\hline
Adjacency Matrix                & ${\tilde T_{AP}}$     & $N_c$ & $N_r$  \\ \hline
 real system Clements   & 0.16                  & 0.61  & 0.49   \\ \hline
 real system Robertson          & 0.06                  & 0.65  & 0.58   \\ \hline
 real system Kato               & 0.11                  & 0.75  & 0.57   \\ \hline

\end{tabular}
\end{center}

\vspace{1cm}

\caption{\footnotesize $\tilde T_{AP}$ and nestedness coefficient for real  matrices~\cite{clements}
~\cite{Robertson}~\cite{Kato}.} 
\label{T&N}
\end{table}

\section{Conclusions}
We introduce here the nestedness coefficient for plants and animals,  as  an alternative way to measure the 
nestedness of the system. Unlike the Atmar and Patterson "temperature", which measures an 
average distance to a supposed perfectly nested state,  the nestedness coefficient
  is operationally
 defined, based on the robustness of the network. In addition it has the advantage to  easily account for
 the different degrees of order of each guild at any given step of their ordering process.

\end{document}